# A model for confined Tamm plasmon devices

**MIKE ADAMS**[1*]**, BEN CEMLYN**[1]**, IAN HENNING**[1]**, MATTHEW PARKER**[2]**, EDMUND HARBORD**[2] **AND RUTH OULTON**[2]

[1]*School of Computer Science and Electronic Engineering, University of Essex, Wivenhoe Park, Colchester, CO4 3SQ, UK*
[2]*Quantum Engineering Technology Labs, School of Physics and Department of Electrical and Electronic Engineering, University of Bristol, 5 Tyndall Ave, Bristol BS8 1FD, UK*
[*]*adamm@essex.ac.uk*

**Abstract:** It is shown that cavities formed between a multilayer quarter-wave Bragg reflector and a metal mirror which support Tamm plasmons can be modelled by using a hard-mirror approximation including appropriate penetration depths into the mirrors. Results from this model are in excellent agreement with those found by numerical methods. In addition Tamm modes that are laterally confined by the presence of a metallic disc deposited on the Bragg reflector can be described by the effective index model that is commonly used for vertical-cavity surface-emitting lasers (VCSELs). This enables the lateral modes confined by a circular disc to be found from conventional weakly-guiding waveguide theory similar to that used for optical fibres. The resonant wavelengths of these linearly-polarised (LP) guided modes are calculated as functions of disc diameter and other parameters.

## 1. Introduction

Electromagnetic modes bounded by a distributed Bragg reflector (DBR) and a metal mirror are termed Tamm plasmon-polaritons [1], or simply Tamm plasmons, by analogy to the electronic states that can appear at the surface of a crystal [2]. In the present contribution, the term "Tamm modes" will be used. Confined Tamm modes have been demonstrated with lateral confinement limited to the size of a metallic disc deposited on the DBR [3]. Lasing has been reported with Tamm modes for both large-area metal layers [4] and with confinement provided by micron-scale metal discs [5]. Replacing the discs by microrectangles with an aspect ratio of 2 has resulted in polarisation-controlled Tamm lasers [6]. Tamm single-photon sources have been demonstrated with InGaAs/GaAs quantum dots (QDs) emitting at 910 nm [7] and InP/GaInP QDs emitting at 656 nm [8]. Other proposed device applications of Tamm modes include all-optical bistable logic [9], multichannel filters [10] and novel forms of sensors [11,12].

Analysis of one-dimensional Tamm modes is conventionally performed using the transfer matrix method (TMM, see, for example [1,4,9,12-14]). For three-dimensional confined Tamm modes, the modelling has used coupled wave analysis [5], the aperiodic Fourier modal method (a-FMM) [6], and the finite-difference time-domain (FDTD) method [3,14]. All these methods involve numerical computation and do not afford direct physical insight into modal behaviour or identification of trends with variation of structural and material parameters. In the present contribution, drawing on concepts well-established for modelling vertical-cavity surface-emitting lasers (VCSELs) and resonant-cavity light-emitting diodes (RCLEDs), we offer a novel approach to modelling one-dimensional and three-dimensional Tamm modes that includes all the relevant physics and offers a simpler approach to studying trends.

In the next section the hard-mirror model for a one-dimensional cavity is applied to study Tamm modes with the aid of the concept of the penetration depth of the field into the DBR [15]. The following section extends this approach to describe three-dimensional confined Tamm plasmon modes by using the effective index model [16]. After that, a numerical example is given and compared with results from the literature where numerical methods have been used, and finally we summarise our conclusions and the outlook for further applications of our approach.

## 2. One-dimensional approximation

A general one-dimensional structure to support Tamm modes is shown schematically in Fig. 1(a). It consists of a spacer layer of thickness $L_S$ and refractive index $n_S$, between a multilayer quarter-wave Bragg reflector and a metal whose complex refractive index is $(n_M + ik_M)$. Fig 1(b) shows the hard-mirror model for this cavity, allowing for penetration of the field into the DBR and the metal.

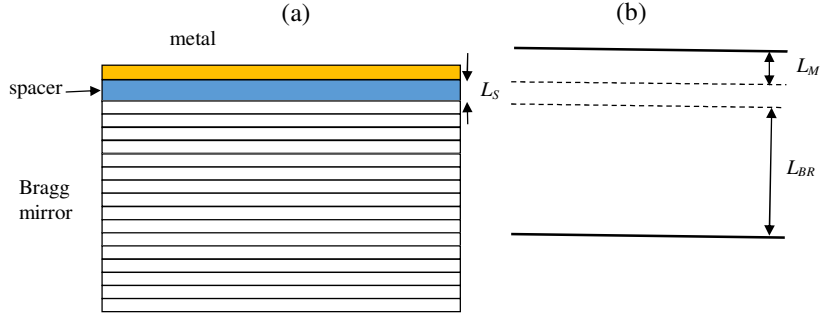

Fig. 1 (a) Schematic of one-dimensional structure to support Tamm modes. (b) Equivalent hard-mirror model.

From standard VCSEL and RCLED cavity modelling (see, e.g., [17,18]), we approximate the confinement of the Tamm mode between the DBR and the metal layer as a cavity formed by two hard mirrors each of which is positioned at a penetration depth into the appropriate medium. The penetration depths are denoted by $L_M$ and $L_{BR}$, respectively, for the metal and Bragg reflector, as shown in Fig 1(b). Rigorous expressions for the penetration depth $L_{BR}$ are derived in [15] in terms of refractive indices, number of layers in the DBR and wavelength. Simpler analytical expressions that are very accurate for high numbers of layers when the penetration length saturates are given in [19]. The penetration depth into the metal, $L_M$, at the centre (Bragg) angular frequency $\omega_0$ of the DBR can be calculated from the phase change, $\beta$, on reflection [20]:

$$L_M = \frac{c}{2\omega_0 n_M}(\pi - \beta) \qquad (1)$$

where $c$ is the speed of light. An explicit (but lengthy) expression for $\beta$ is given in [21] in terms of refractive indices, thickness of the metal and wavelength. In the limit of large layer thickness this expression reduces to the simple form obtained from the phase change on reflection from an isolated metal surface [22]. Usually $\beta$ lies in the second quadrant.

The longitudinal modes of the structure occur at frequencies for which the round-trip phase is an integer multiple of $2\pi$. Allowing also for the reflection delay $\tau$ from the Bragg mirror [15], the sum of phase terms in the cavity round-trip gives

$$2\frac{\omega_N}{c}(n_S L_S + \bar{n} L_{BR}) + 2\frac{\omega_0}{c} n_M L_M - \omega_0 \tau = (2N-1)\pi \qquad (2)$$

where $\omega_N$ is the resonant angular frequency of mode $N$ (=1,2,3,…) and the average refractive index in the Bragg mirror is defined [19] as

$$\bar{n} = \frac{n_H d_H + n_L d_L}{d_H + d_L} \qquad (3)$$

with $n_{H,L}$ and $d_{H,L}$ as the refractive indices and thicknesses of the high and low index layers in the Bragg mirror. The reason for the term $(2N-1)$ in equation (2) is that the combination of the spacer layer and the metal thickness should behave as a quarter-wave layer [23].

The delay time $\tau$ is related to the penetration depth in the DBR by

$$\tau = \frac{2\bar{n}}{c} L_{BR} \qquad (4)$$

Using (4), equation (2) can be re-written in the form

$$\frac{\omega_N}{c} n_S L_S + \frac{\omega_0}{c} n_M L_M + \frac{(\omega_N - \omega_0)}{c} \bar{n} L_{BR} = \left(N - \frac{1}{2}\right)\pi \qquad (5)$$

Equation (5) is the cavity resonance condition in terms of the phase changes in the spacer and on reflection from the metal and the DBR. It shows explicitly that when the mode frequency occurs at the centre frequency of the grating, there is no dependence on the penetration depth $L_{BR}$. This is because in this case the phase on reflection from the grating is either zero or $\pi$. Equation (5) here is equivalent to equation (5) in [1], with the difference that in [1] the refractive index of the metal is given by the Drude model and there is no spacer layer ($L_S = 0$).

From (5) it follows that the mode frequency $\omega_N$ is given by

$$\omega_N = \frac{\omega_0 (\bar{n} L_{BR} - n_M L_M) + c\pi\left(N - \frac{1}{2}\right)}{n_S L_S + \bar{n} L_{BR}} \qquad (6)$$

For the calculation of $L_{BR}$, we use the phase penetration depth $L_\tau$ [15]:

$$L_\tau = \frac{\pi c}{2\omega_0 \bar{n}} \frac{q}{1-p} \frac{(1 - a^2 p^{m-1})(1 - p^m)}{(1 - q^2 a^2 p^{2m-2})} \qquad (7)$$

where $m$ is the number of layers in the DBR and $a, p, q$ are refractive index ratios (low to high) of the three types of interfaces that characterise the mirror:

$$a = \frac{n_{LE}}{n_{HE}} \qquad p = \frac{n_L}{n_H} \qquad q = \frac{n_{LI}}{n_{HI}} \qquad (8)$$

The ratio $a$ applies to the interface between the last DBR layer and the exit medium (subscript $E$) and the ratio $q$ refers to the interface between the incident medium (subscript $I$) and the first DBR layer. We note here that the simpler expressions for penetration depth [15,19] are not sufficiently accurate for the numerical example considered below because the index differences are too large.

In the numerical work below the full expression [21] for the phase $\beta$ will be used in equation (1) for the penetration depth of the metal in order to include the effect of a finite layer.

## 3. Three-dimensional approximation

To model confinement in the lateral direction, we follow the effective index approach proposed in [16] and now widely used for VCSEL design. The essence of this method is that the lateral effective index profile is determined by the local changes of the Fabry-Perot cavity resonance wavelength. So for a circular metal disc on top of the vertical structure discussed above, the change of effective index, $\Delta n_e$, is related to the change of resonant wavelength, $\Delta\lambda$, by

$$\frac{\Delta n_e}{n_e} = \frac{\Delta \lambda}{\lambda} \qquad (9)$$

where $n_e$ is the effective index of the cavity given by the mean value of indices in the spacer and the Bragg mirror:

$$n_e L_{eff} = n_S L_S + \bar{n} L_{BR} \qquad (10)$$

with the total effective cavity length, $L_{eff}$, defined as

$$L_{eff} = L_S + L_{BR} \qquad (11)$$

The resulting radial index profile is illustrated in Fig. 2(a), with a schematic of the physical structure in Fig 2(b).

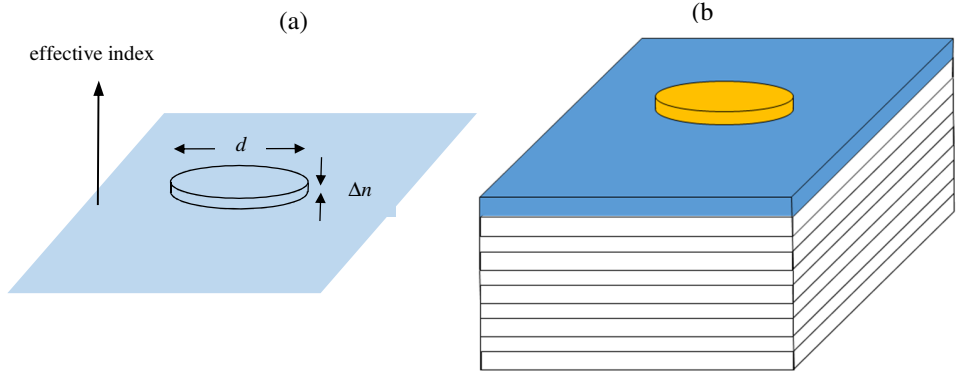

Fig. 2 (a) Effective index model of circularly-symmetric cylindrical waveguide for the three-dimensional structure to support confined Tamm modes shown in (b).

For the region outside that beneath the metal disc, the upper mirror of the effective Fabry-Perot cavity is formed by the semiconductor-air interface. Hence in this region, equation (5) is replaced by

$$\frac{\omega_{Nc}}{c} n_S L_S + \frac{(\omega_{Nc} - \omega_0)}{c} \bar{n} L_{BR} = \pi \left( N - \frac{1}{2} \right) \quad (12)$$

where $\omega_{Nc}$ is the mode frequency in this region. It follows that the quantity $\Delta\lambda$, required in equation (9) is given by

$$\Delta\lambda = 2\pi c \left( \frac{1}{\omega_N} - \frac{1}{\omega_{Nc}} \right) \quad (13)$$

Thus solutions of equation (5) and (12) can be used in (13) and (9) to calculate the effective index profile in the radial direction.

The resulting circular waveguide of diameter $d$, core index $n_e$, and core-cladding index difference $\Delta n_e$ ($\ll n_e$) supports modes that are, to a very good approximation, linearly-polarised (LP) by analogy with the modes of weakly-guiding optical fibres [24]. These LP modes are characterised by the index eigenvalues, $n$, of the wave equation such that their cut-off frequency is given by $n = n_e - \Delta n_e$. These modes are characterised by 2 subscripts, one for each of the radial and azimuthal co-ordinates. We will use $l$ as the azimuthal subscript and $p$ as the radial subscript, so that the mode is labelled $LP_{lp}$. Mode solver software for the LP modes is widely available for optical fibres and can simply be used for the corresponding confined Tamm modes with appropriate definitions of core and cladding indices as described above.

Assuming the solutions for $n$ have been found, the frequency of the confined mode, $\omega_p$, can be calculated by using again the cavity resonance condition in the form:

$$\omega_{lp} = \frac{\omega_0 (\bar{n} L_{BR} - n_M L_M) + c\pi \left( N - \frac{1}{2} \right)}{n L_{eff}} \quad (14)$$

Equation (14) indicates that the confined Tamm resonant wavelength is directly proportional to the index eigenvalue for the corresponding LP mode.

## 4. Numerical example

The method outlined above will be illustrated by applying it to the structure designed for Tamm modes in [14], which is the only publication (to the best of our knowledge) that does not use the Drude model for the metal layer and is thus closest to our treatment here. The structure considered in [14] has a GaAs spacer, an AlAs/GaAs DBR with 35 layers, and a GaAs substrate. Hence the incident and exit media have the same (high) refractive index [25], and the first and last layers of the DBR have the same (low) index [26]. The metal is gold with $n_M = 0.38$ and $k_M = 8.7$ in the wavelength range of interest [27]. Numerical values of the semiconductor parameters are given in Table 1. The stopband of the Bragg mirror lies between 1205 and 1375 nm [14], so this defines the value of the Bragg frequency.

**Table 1. Semiconductor parameter values**

| | |
|---|---|
| $\lambda_0$ (nm) | 1284.4 |
| $\omega_0$ (x $10^{15}$ rad/s) | 1.4676 |
| $n_S, n_H, n_{Hl}, n_{HE}$ | 3.409 |
| $n_L, n_{Ll}, n_{LE}$ | 2.910 |
| $p,q,a$ | 0.854 |
| $d_H$ (nm) | 94.2 |
| $d_L$ (nm) | 110.4 |
| $\bar{n}$ | 3.139 |
| $L_{DBR}$ (nm) | 591.9 |

Fig. 3 shows the variation of phase with thickness of the gold layer calculated using the expression in [21].

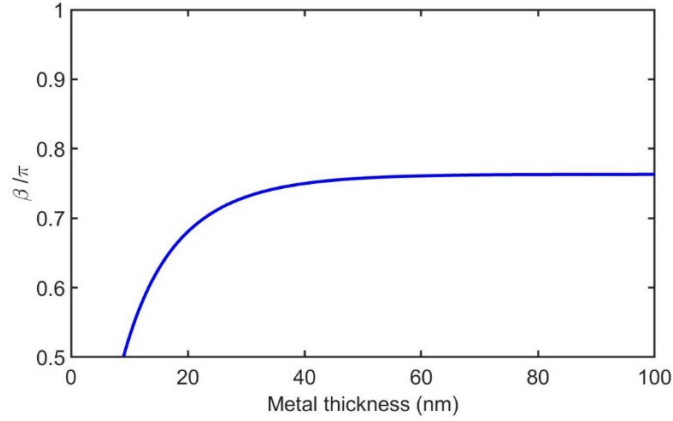

Fig. 3 Phase on reflection from metal (gold) versus thickness

Using equation (6) with N = 1 for a 25 nm gold thickness and 75 nm spacer gives $\omega_1 = 1.4487$ x $10^{15}$ rad/s, corresponding to a Tamm wavelength of 1301.1 nm. This compares well with the value of 1304 nm reported for this structure in [14] where the TMM was used. It is interesting also to estimate the Tamm resonance from the approximation used in [1] where the Drude model for the metal is used. Adapting the result from [1] to allow for the presence of a spacer layer gives

$$\omega_1 = \frac{\omega_0 \left(1 + \dfrac{\Delta n}{2 n_H}\right)}{1 + \dfrac{\omega_0 \Delta n}{\pi}\left(\dfrac{2}{\omega_p \sqrt{\varepsilon_b}} + \dfrac{n_S L_S}{n_H c}\right)} \qquad (15)$$

where $\Delta n = n_H - n_L$, $\omega_p$ is the plasma frequency and $\varepsilon_b$ is the background dielectric constant. Using our parameter values in (15) with a plasma energy of 8.9 eV [1] yields a frequency of 1.441 x 10$^{15}$ rad/s, corresponding to a Tamm wavelength of 1308 nm. This level of agreement may be fortuitous, since the approximation in (15) takes no account of the gold thickness nor of the number of layers in the Bragg mirror, effectively assuming both these quantities are approaching infinity. Since the penetration depth into the DBR is approaching saturation for 35 layers we do not expect additional layers to affect the resonant wavelength from (6) significantly. However, it is clear from Fig. 3 that the phase on reflection from the mirror is some way from saturation at a thickness of 25 nm and hence a change of the Tamm wavelength is expected, as we demonstrate below.

Fig. 4 shows the Tamm resonance wavelength as a function of (a) spacer thickness with the gold thickness constant at 25 nm, and (b) gold thickness with the spacer thickness kept constant at 75 nm. Again these results are in excellent agreement with those in Fig. 2 of [14].

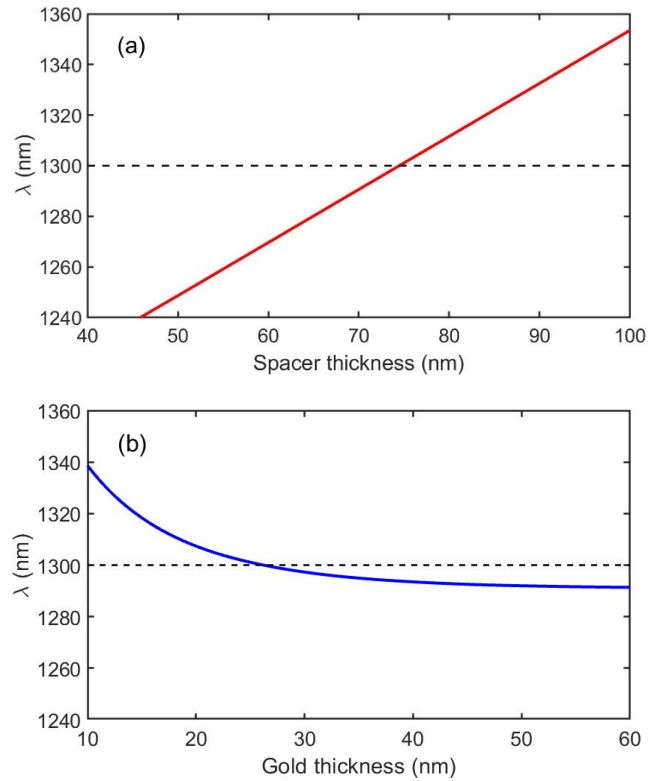

Fig. 4 Calculated Tamm wavelength (a) versus spacer thickness for fixed gold thickness of 25 nm, and (b) versus gold thickness for fixed spacer thickness for 75 nm.

Turning to the two-dimensional model, for the case of a 75 nm spacer and 25 nm of gold, in the region outside that beneath the metal disc equation (12) is used with N = 1 to calculate a wavelength of 1245.8 nm. Hence we can use equation (9) with $\lambda$ = 1307.7 nm, $\Delta\lambda$ = 55.3 nm and $n_e$ = 3.170 (from equation (10)) to calculate the value of $\Delta n_e$ as 0.135. For these parameter values, the cut-off values ($d_c$) of the disc diameter for the LP modes can be calculated. Table 2 gives the values of $d_c$ for some lower-order LP modes using the effective index and index difference values calculated for this example. Each LP mode corresponds to hybrid (HE or EH), transverse electric (TE), or transverse magnetic (TM) modes, as indicated in Table 2. The number of lateral modes increases rapidly with disc diameter. Single-mode operation is only found for a diameter less than 1.1 μm.

Table 2. Cut-offs of LP modes.

| LP mode | Equivalent modes | $d_c$ (μm) |
|---|---|---|
| $LP_{01}$ | $HE_{11}$ | 0 |
| $LP_{11}$ | $HE_{21},TE_{01},TM_{01}$ | 1.10 |
| $LP_{02},LP_{21}$ | $HE_{12},EH_{11},HE_{31}$ | 1.72 |
| $LP_{31}$ | $EH_{21},HE_{41}$ | 2.30 |
| $LP_{12}$ | $HE_{22},TE_{02},TM_{02}$ | 2.47 |
| $LP_{41}$ | $EH_{41},HE_{51}$ | 2.86 |
| $LP_{03},LP_{22}$ | $HE_{13},EH_{12},HE_{32}$ | 3.14 |
| $LP_{51}$ | $EH_{41},HE_{61}$ | 3.40 |
| $LP_{32}$ | $EH_{23},HE_{42}$ | 3.77 |
| $LP_{13}$ | $HE_{23},TE_{03},TM_{03}$ | 3.88 |
| $LP_{61}$ | $EH_{51},HE_{71}$ | 3.93 |

The calculated variation of resonant wavelength with disc diameter, using equation (14), for the first twelve LP modes is shown in the Fig. 5. The trends here are similar to those measured and calculated in [3] (albeit expressed there in terms of photon energy rather than wavelength) for a somewhat similar structure.

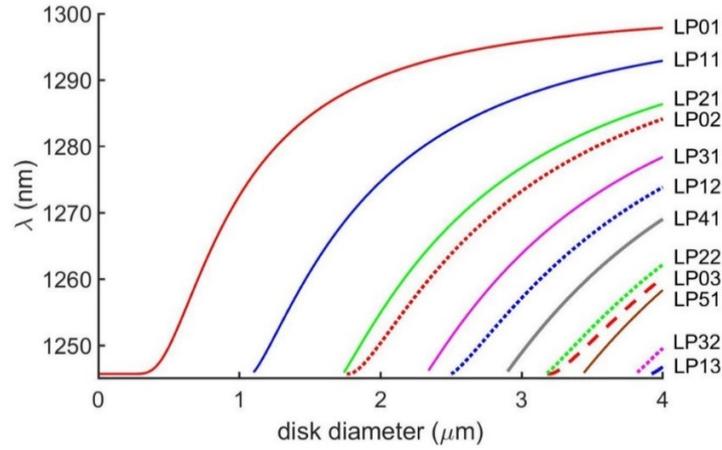

Fig. 5 Calculated Tamm wavelength for the first twelve LP modes versus disc diameter for gold thickness 25 nm and spacer thickness 75 nm.

## 5. Conclusion

We have shown that concepts widely used in design and analysis of VCSELs and RCLEDs can be successfully applied to model Tamm modes confined between a metal layer and a Bragg reflector. The hard mirror approach where an effective cavity is defined to include penetration depths of the optical wave into the metal and the Bragg mirror successfully predicts the resonant wavelengths of one-dimensional Tamm modes. It is worth noting that this approach uses the calculated phase on reflection from the metal, using measured real and imaginary parts of the refractive index rather than invoking the Drude model. Three-dimensional Tamm modes can be modelled using the effective index approach where the radial effective index profile is found from the difference in resonant wavelengths between the one-dimensional mode beneath a circular metal disc and that outside this region. This enables the number of transverse modes and their wavelengths to be determined as functions of disc diameter and other parameters, a task which is not straightforward with purely numerical computational approaches.

It is clear that the methods proposed here can be applied to a wider range of structures than those currently of interest. Whilst attention has been confined here to circular discs, the method proposed could just as well be applied to discs of other shapes, e.g. elliptical, racetrack or rectangular, since the basic principle of decoupling the vertical field component from the lateral components applies independently of the geometry. Thus, for example, the method could be used to study polarisation properties of non-circularly symmetric discs [6] or more complicated shapes of the metal layer. Further developments could include analysis of the effects of metallic layers and/or spacers of differing thicknesses in the lateral directions, thus opening up the prospect of integrating novel Tamm modal devices on a single Bragg mirror in order to produce multifunctional components for a variety of potential applications.

## Funding

UK Engineering and Physical Sciences Research Council (Grant Nos. EP/G012458/1, EP/M024237/1, EP/M024156/1, EP/N003381/1).